\documentclass[11pt]{article}
\usepackage{coling2020}
\usepackage{times}
\usepackage{url}
\usepackage{latexsym}

\usepackage{graphicx}
\usepackage{xcolor}
\usepackage{multirow,rotating}
\usepackage{amsmath,amssymb,bm}
\usepackage{booktabs}
\usepackage{subfig}
\usepackage[textwidth=0.75in,textsize=small
]{todonotes}

\colingfinalcopy 

\title{Embedding Meta-Textual Information for Improved Learning to Rank}

\author{Toshitaka Kuwa$^{*}$, Shigehiko Schamoni$^{\dagger,\ddagger}$, Stefan Riezler$^{\dagger,\ddagger}$ \\
  $^{\dagger}$Department of Computational Linguistics, Heidelberg, Germany. \\
  $^{\ddagger}$Interdisciplinary Center for Scientific Computing (IWR), Heidelberg, Germany \\
  {\tt \{kuwa, schamoni, riezler\}@cl.uni-heidelberg.de} \\}

\date{}

\begin{document}
\maketitle
\vspace{-2em}
\begin{abstract}
  Neural approaches to learning term embeddings have led to improved computation of similarity and ranking in information retrieval (IR). So far neural representation learning has not been extended to meta-textual information that is readily available for many IR tasks, for example, patent classes in prior-art retrieval, topical information in Wikipedia articles, or product categories in e-commerce data. We present a framework that learns embeddings for meta-textual categories, and optimizes a pairwise ranking objective for improved matching based on combined embeddings of textual and meta-textual information. We show considerable gains in an experimental evaluation on cross-lingual retrieval in the Wikipedia domain for three language pairs, and in the Patent domain for one language pair. Our results emphasize that the mode of combining different types of information is crucial for model improvement. 
\end{abstract}

\section{Introduction}
\label{intro}

%
%
\blfootnote{
    %
    %
    %
    %
    %
    %
    \hspace{-0.2cm}
      $^{*}$ This work was done while the author visited the Department of Computational Linguistics, Heidelberg, Germany.\\
     This work is licensed under a Creative Commons 
     Attribution 4.0 International License.
     License details:
     \url{http://creativecommons.org/licenses/by/4.0/}.
}

The recent success of neural methods in IR rests on bridging the gap between query and document vocabulary by learning term embeddings that allow for improved computation of similarity and ranking \cite{MitraCrasswell:18}. So far vector representations in neural IR have been confined to textual information, neglecting information beyond the raw text that is potentially useful for retrieval and is readily available in many data situations. For example, meta-textual information is available in the form of patent classes in prior-art search, topic classes in search over Wikipedia articles, or product classes in retrieval in the e-commerce domain. A straightforward approach to use such meta-textual information to improve retrieval would require an exact match of meta-textual categories of queries and relevant documents. However, in the majority of cases the sets of categories assigned to queries and documents overlap only at a small percentage, for both relevant and irrelevant documents. Thus, more sophisticated techniques are needed to aid similarity computation and ranking by meta-textual category information. 

In this paper, we show how to apply neural embedding methods to meta-textual categories. We show that meta-textual information is an incomplete representation of queries and documents and does not suffice for a standalone computation of similarity and ranking. 
However, an incorporation of pre-trained embeddings of meta-textual categories into a neural learning-to-rank approach yields significant improvements over a learning-to-rank approach that uses text-only embeddings \cite{SasakiETAL:18}. In our approach, enhanced embeddings are created by concatenating text embeddings with meta-textual category embeddings, and a fully connected weight layer is learned on top of the concatenation of the enhanced embeddings of a query-document pair by a deep multilayer perceptron that is optimized for pairwise ranking. We present experiments on three different language pairs for cross-lingual retrieval in the Wikipedia domain, and show improvements of up to 2~NDCG points by incorporating meta-textual embeddings in a learning-to-rank framework. Additional experiments on a single language pair in the Patent domain also show improvements of up to 1.3~NDCG points by incorporating embeddings for patent classifications, and up to 6.3 NDCG points if separate models for text and meta information are combined in an ensemble. 

\section{Related Work}

Embeddings of meta-textual information via so-called ``side constraints'' has been done successfully in neural machine translation applications \cite{ChenETAL:16,SennrichETAL:16,SennrichETAL16linguistic,ChuETAL:17,KobusETAL:17,JohnsonETAL:17,JehlRiezler:18}. Meta-information such as domain labels, product categories, or politeness level, has been incorporated in various ways, including a choice of feeding a meta-information tag at the source side or the target side, the choice of attaching it to each sentence or to each word, and the choice of learning embeddings vi learning-to-rank training or via devising a specific representation for the meta-information embeddings. 

An incorporation of meta-information into learning-to-rank approaches for IR has previously been presented by \newcite{SchamoniRiezler:15}, albeit not in a neural IR framework. 

The neural learning-to-rank architecture used in our work builds on \newcite{SasakiETAL:18}. While they present an option to use learned text embeddings for cosine similarity scoring, our approach focuses on direct learning of a similarity function in order to learn a proper weigthing of textual and meta-textual embeddings.

\section{Learning-to-Rank via Text and Meta-Text Embeddings}

\subsection{Convolutional Embeddings of Textual Information}
\label{sec:text-emb}

Similar to the approach of \newcite{SasakiETAL:18}, that functions as baseline in our work, we employ a neural learning-to-rank model that learns a relevance score $S(\vec{c}_q,\vec{c}_d)$ for a vector representation of an English query $\vec{c}_q$ and a foreign-language document $\vec{c}_d$. These vector representations are computed by a convolutional feature map over a ``sentence matrix'', with rows consisting of vector representations of words in a query or document (pre-trained using word2vec \cite{MikolovETAL:13} on the corpora described below), and columns in the size of the length of the query or document. This choice of embeddings might seem simplistic compared to recent approaches to contextual word embeddings \cite{DevlinETAL:2018,PetersETAL:2018}, however, it is motivated by the goal of understanding the relative benefit of meta-text embeddings based on a manageable architecture for learning-to-rank with text-only embeddings.

Let $\vec{x}_{1:n} = [\vec{x}_1; \vec{x}_2;\ldots;\vec{x}_n]$ be the concatenation of word vectors for a query or a document of $n$ words, where each word vector is of dimensionality $k$, and let $\vec{x}_{i:i+h-1} = [\vec{x}_i; \vec{x}_{i+1};\ldots;\vec{x}_{i+h-1}]$ denote the concatenation of word vectors in a window of width $h$ starting from position $i$. The parameters of a convolution involve a filter $\vec{W} \in \mathbb{R}^{hk \times m}$ which is applied to a window of $h$ words, and extracts vectors $\vec{p}_i, i= 1, \ldots, l$ where 
\[\vec{p}_i = f(\vec{x}_{i:i+h-1} \cdot \vec{W}+b)\] 
and $\vec{p}_i \in \mathbb{R}^m$, $b \in \mathbb{R}^m$ and $f$ is a non-linear function such as hyperbolic tangent. The final feature representation extracts an $m$-dimensional vector $\vec{c}$ by average pooling over time where $\vec{c} = \mathrm{avg}_{1 < i \leq l} \vec{p}_i$. Applying this procedure to ``sentence matrices'' consisting of concatenations of word representations of queries or documents yields our representations $\vec{c}_q$ and  $\vec{c}_d$.

A further step involves learning a multi-layer perceptron with a fully connected layer on top of the concatenation $[\vec{c}_q;\vec{c}_d]$, defining a relevance score 
\[S(\vec{c}_q,\vec{c}_d) = \tanh(\vec{O} \cdot \mathrm{relu}(\vec{U} \cdot [\vec{c}_q;\vec{c}_d]^\top)),\]
where $\vec{O} \in \mathbb{R}^{1 \times s}$, $\vec{U} \in \mathbb{R}^{s \times 2m}$, and $s$ is the dimensionsonality of the hidden state.

\begin{figure}[t!]
\centering
\includegraphics[width=0.5\textwidth]{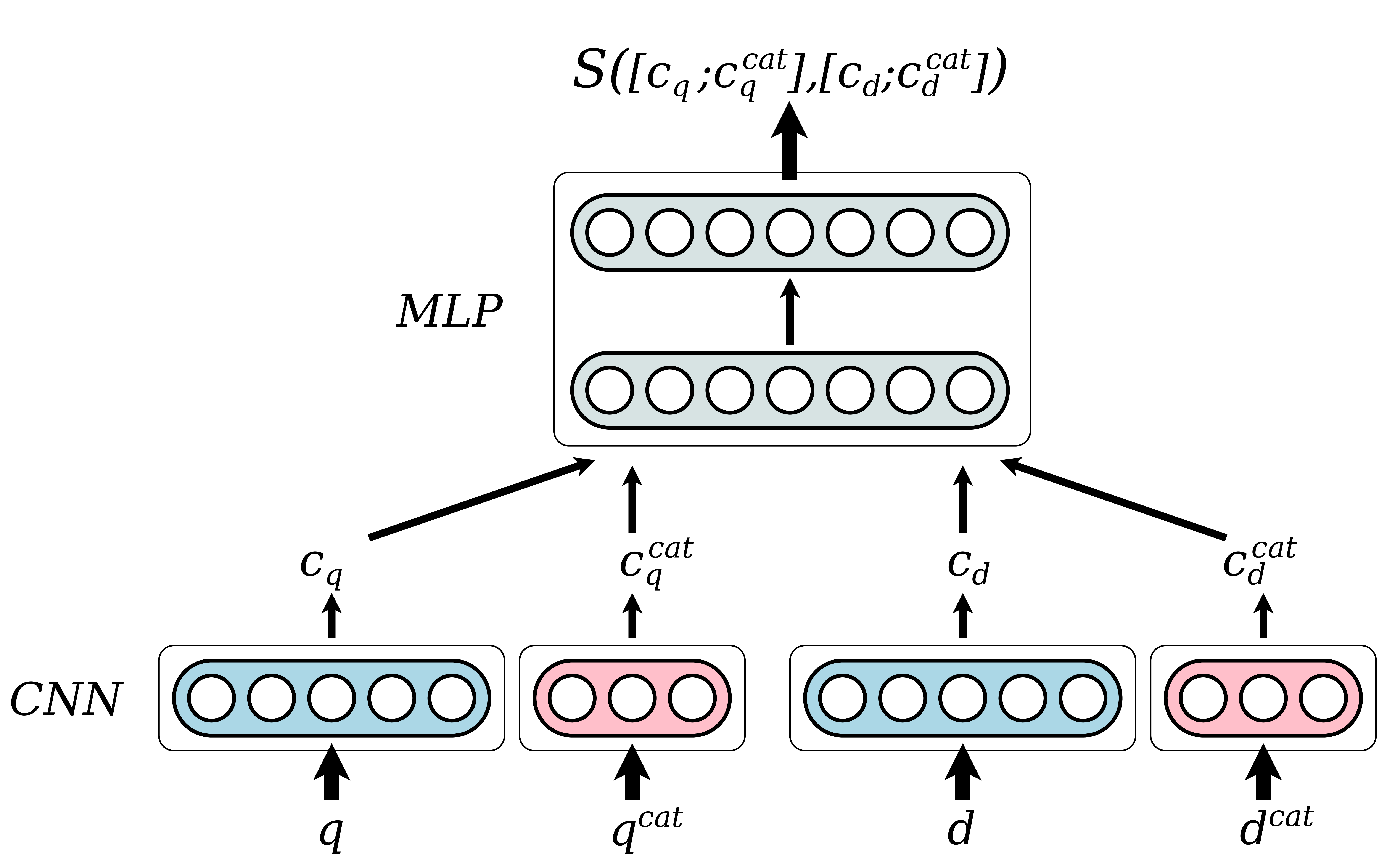}
\vspace{-2mm}
\caption{Neural learning-to-rank architecture with embeddings of text ($q$,$d$) and categories ($q^{cat}$,$d^{cat}$) of queries and documents. CNN components generate compressed representations of texts ($c_q$,$c_d$) and categories ($c_q^{cat}$,$c_d^{cat}$), the MLP component learns a relevance score $S$ between queries and documents.} \label{fig1} 
\end{figure}

Using the shorthand $\vec{\Theta} = \{\vec{W}, b, \vec{O}, \vec{U}\}$ for the parameters to be learned, we can write the function to be optimized as the pairwise ranking objective 
\[
L = \max_{\vec{\Theta}} \{0, 1- \left( S(\vec{c}_q,\vec{c}_{d^+}) - S(\vec{c}_q,\vec{c}_{d^-})\right)\},
\]
where $d^+$ and $d^-$ are relevant and non-relevant document respectively. During training, the model learns the convolutional filters and the similarity function. An overview of the overall network architecture is given in Figure \ref{fig1}.

\subsection{Incorporating Convolutional Embeddings of Meta-Textual Data}
\label{sec:category-emb}

To enrich the representation of queries and documents, we further generate dense vectors, i.e. embeddings, for meta-textual information. In the case of Wikipedia, we make use the category graph to extract embeddings that potentially add useful information to the retrieval task. 
We opt for a machine learning method to generate graph embeddings for each node in our category graph. From the several available methods for learning graph embeddings we follow the DeepWalk-idea described by \newcite{PerozziETAL:14}. Our DeepWalk-based strategy consists of 3 steps: 
\begin{enumerate}
    \item For all categories in the category graph, apply a random walk of predefined length to generate context sequences. 
    \item Apply word2vec's \cite{MikolovETAL:13} skip-gram with negative sampling method to learn a model that predicts the context given a category. 
    \item Use the trained model to calculate an embedding for each category.
\end{enumerate}

More computationally advanced algorithms such as GraphGAN \cite{FigueiredoETAL:2017} or struc2vec \cite{WangETAL:2017} show better performance on the Wikipedia node classification task, however, the improvements compared to DeepWalk are small: 1--2\% in accuracy, 0.01 in Macro-F1.\footnote{\url{https://paperswithcode.com/sota/node-classification-on-wikipedia}, accessed 06/08/2020.} 
As a matter of fact, the Wikipedia category graph is noisy and contains oddities like loops, or missing and wrong connections. We thus chose this method for its simplicity and robustness. 
We did not apply any weighting of node connections as suggested in Node2vec \cite{GroverLeskovec:2016}. 
A single random walk of length 40 was generated for each category instance. All hyperparameters were tuned on the dev set. 

Each query's and document's category information is now encoded as a set of embeddings. 
Typically, the number of elements in these sets is much smaller than the words in the associated texts, but we nevertheless reused the convolutional architecture for words to get compressed category representations. The concatenation of $n$ categories is then processed via convolutions 
$\vec{x}^{cat}_{1:n} = [\vec{x}^{cat}_1; \vec{x}^{cat}_2;\ldots;\vec{x}^{cat}_n]$. 
In contrast to processing of word sequences, we eliminate order specific information for categories by fixing the window size to $h=1$, effectively treating the ordered category sequence as an unordered set. Training a convolution $\vec{p}^{cat}_i = f(\vec{x}^{cat}_{i} \cdot \vec{W}^{cat}+b^{cat})$
followed by average pooling yields our compressed category representations $\vec{c}^{cat}_q$ and $\vec{c}^{cat}_d$. 

As before, we fix the embeddings and train only the convolutional parameters and the similarity function $S$ to obtain meaningful representations of categories for the given queries and documents.

\section{Data Sets}

For our CLIR experiments, we conduct experiments on two different domains, namely Wikipedia articles and patents. For the Wikipedia retrieval task, we extend data from  \newcite{SasakiETAL:18} with category information from Wikipedia. For the patent retrieval task, we extend data previously published by \newcite{SokolovETAL:2013b} with information from the International Patent Classification (IPC) system. Table~\ref{table_statistics} lists the statistics of the extended datasets. Both data extensions are publicly available.\footnote{\url{https://www.cl.uni-heidelberg.de/metaclir/}}

\begin{table}[t]
\centering
\caption{Statistics of our two datasets. Listed are the total number of target documents and their meta types in the corpora, and the number of source queries for train, dev, and test sets. For Wikipedia, meta types are the unique categories, for patents they are the unique IPC-codes. } 
\label{table_statistics}
{\footnotesize
\begin{tabular}{ccccccccc}
\toprule
dataset & language & \#~target & \#~meta types & \#~train & \#~dev &  \#~dev & \#~test & \#~test \\
        & pair     & documents & source/target & queries & queries & documents & queries & documents \\
\midrule
 & En-Ja & $ 1,071,292 $ & $ 1,023\text{M}/172\text{k}$ & $298,468 $ & $ 42,638$ & \textit{same} & $ 42,638$ & \textit{same} \\
\textbf{Wikipedia}  & En-De & $ 2,091,278 $ & $ 1,023\text{M}/282\text{k} $ & $ 656,735 $ & $ 93,820 $ & \textit{as} & $ 93,819 $ & \textit{as} \\
 & En-Fr & $ 1,894,397 $ & $ 1,023\text{M}/332\text{k} $ & $ 762,336 $ & $ 108,905 $ & \textit{target} & $ 108,905 $ & \textit{target} \\
\midrule
\textbf{Patent} & Ja-En & $ 1,039,656 $ & $ 23\text{k}/50\text{k} $ & $ 107,061 $ & $ 2,000 $ & $100,000$ & $ 2,000 $ & $100,000$\\
\bottomrule
\end{tabular}
}
\end{table}

\subsection{Cross-Lingual Retrieval in Wikipedia}

The task of cross-lingual retrieval on Wikipedia data is as follows: given a query in a source language, identify the corresponding article in the target language and all other articles that are relevant to this target article, i.e. having incoming and outgoing links to the article in the target language. Pages that connect articles which are irrelevant to each other, e.g. disambiguation pages, were removed from the dataset. Removal of seemingly useless categories such as stub, tracking, or maintenance categories had no effect on the retrieval performance in the English-Japanese case, thus we did not apply any category filtering on any language pair in our final experiments.

\subsubsection{Multiple Relevance Levels}

The textual data we use is taken from \newcite{SasakiETAL:18}. In their approach, they explicitly consider only the highly relevant articles, i.e. articles that have a direct inter-language link (relevance level $r=2$) in Wikipedia. However, the data set they published contains additional relevance judgements for articles that are considered lesser relevant, i.e. articles that are linked from the most relevant article and contain a link referring back to that article (relevance level $r=1$).
Our approach makes use of this additional information by extending the learning-to-rank strategy to not only compare relevant ($d^+$) and irrelevant ($d^-$) documents, but to include a comparison between highly relevant ($d^{r=2}$) and lesser relevant ($d^{r=1}$) documents. 
To keep the data size manageable during training, we sample a subset of pairs from the set of all possible pairings. 

\subsubsection{Category Overlap}
\label{sec:category_overlap}

To estimate the effectiveness of the random walk strategy for generating category embeddings and to investigate the overall utility of such embedded categories, we computed the \textit{category overlap} between queries and relevant documents. This number is calculated as the intersection of translated query categories (using the Wikipedia interlanguage-links) and the categories of the candidate document. 

\begin{figure}[t]
\centering
\subfloat[English-Japanese]{\includegraphics[width=.31\linewidth]{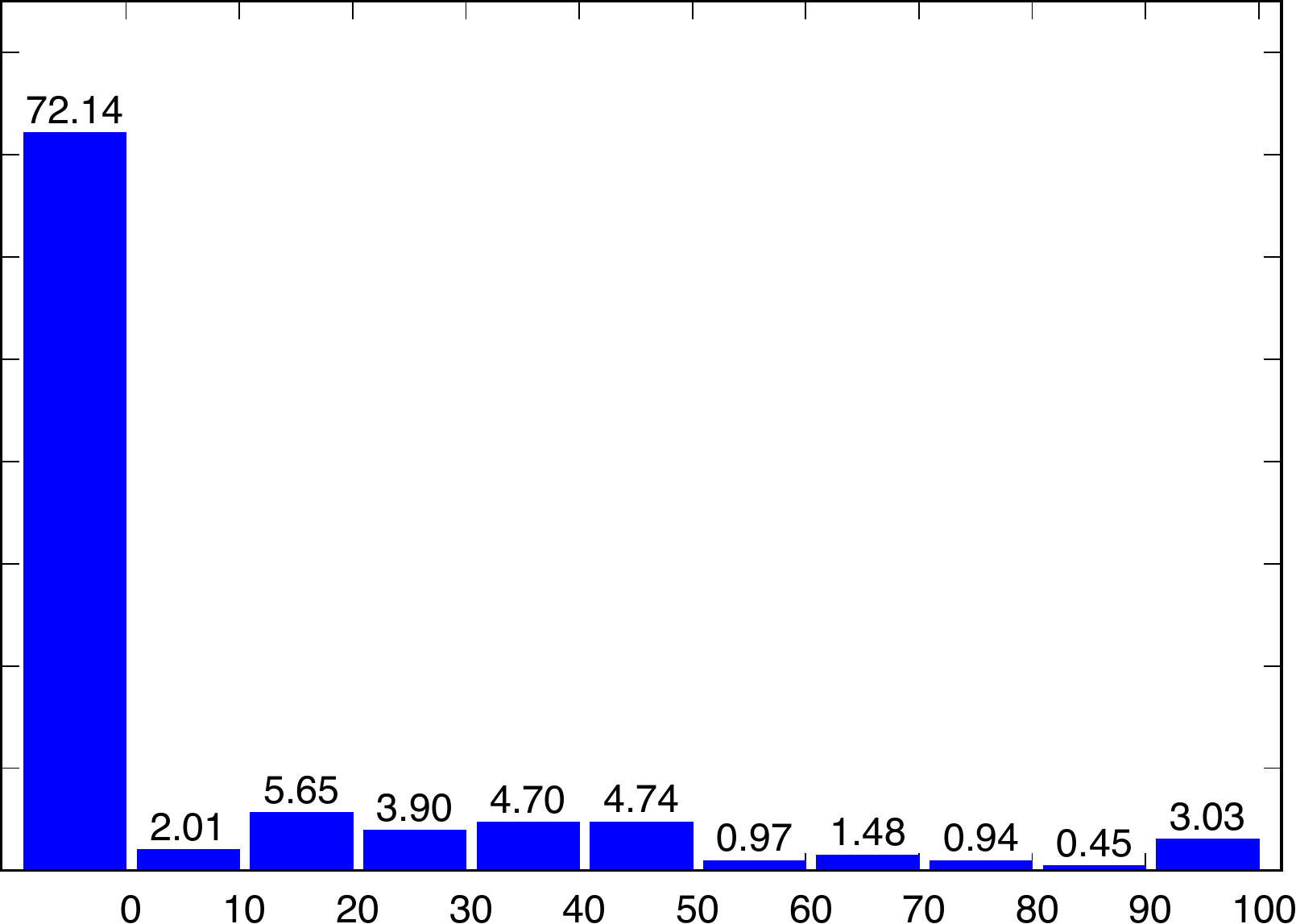}}\quad
\subfloat[English-German]{\includegraphics[width=.31\linewidth]{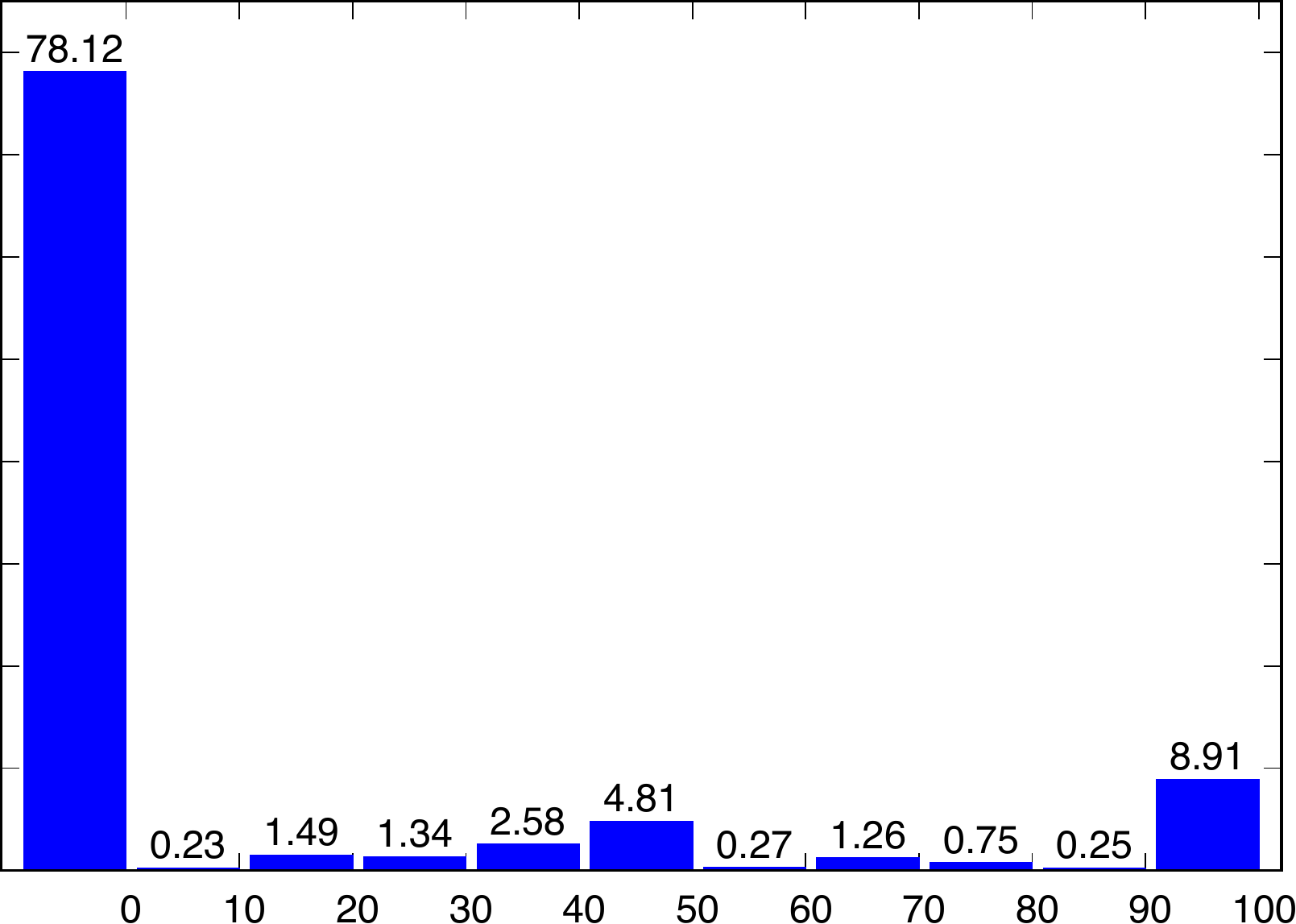}}\quad
\subfloat[English-French]{\includegraphics[width=.31\linewidth]{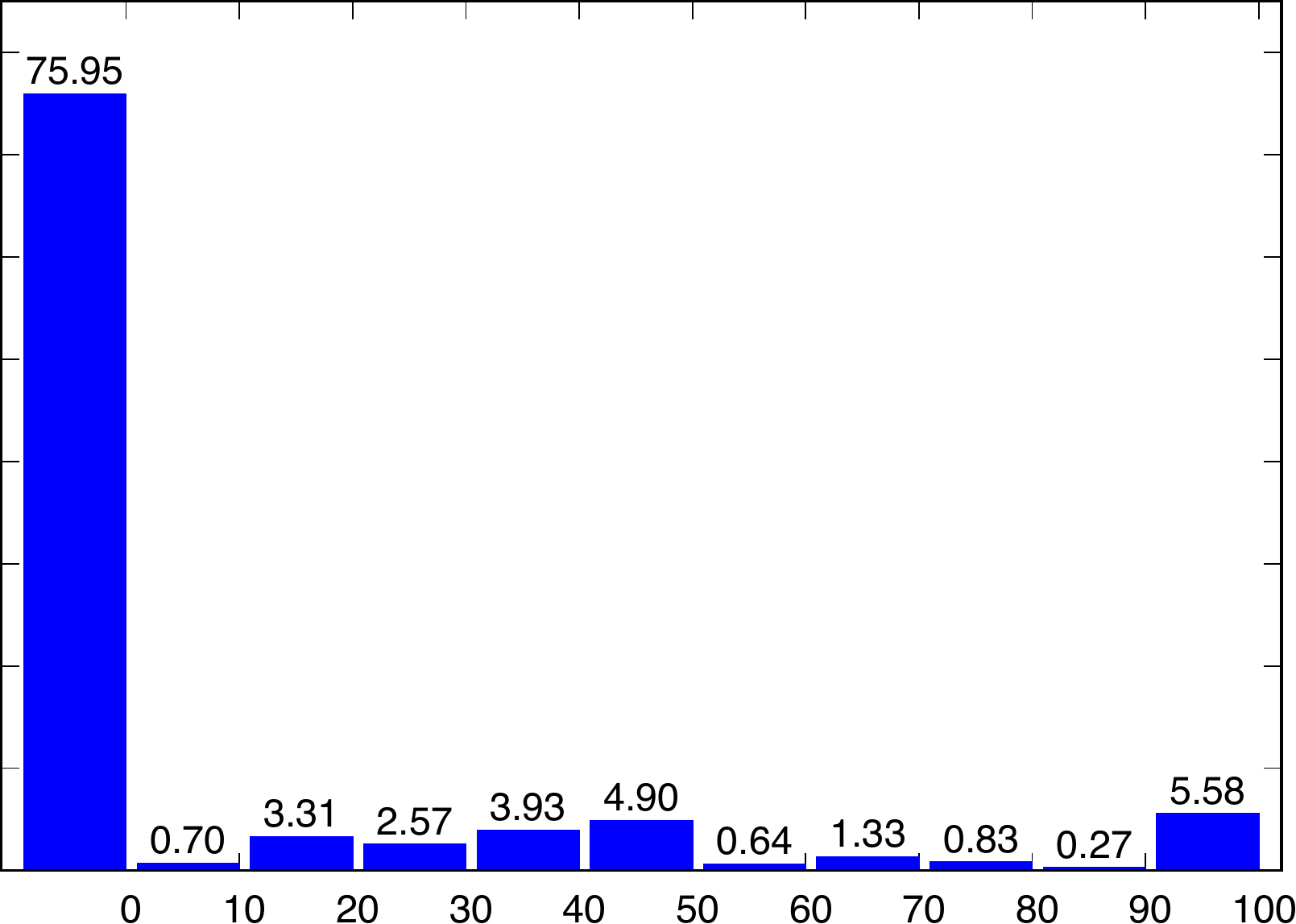}}%
\caption{Percentage of common categories between English queries and relevant documents in Japanese~(a), German (b), and French (c), aggregated into 10\%-buckets. The bar left of 0 denotes the percentage of query-document pairs without a single common category. The category overlap is calculated on the document's side.}
\label{fig:category_overlap}
\end{figure}

As shown in Figure \ref{fig:category_overlap}, the category overlap aggregated over documents of relevance levels $r=1$ and $r=2$ is 0 for over 70\% of cases across all language pairs. A deeper analysis shows that the category overlap between a query and a highly relevant document ($r=2$) is in general higher. For example, on the English-Japanese  retrieval task,  we observe about 14\% of all query-document pairs with a category match between 90--100\%.  However, in 43\% percent of the cases, there is no category overlap even for highly relevant documents. 
The majority of cases with 0 category overlap is due to the overlap between a query and its lesser relevant documents ($r=1$). For English-Japanese, no overlap is found in 76\% of the cases. For the English-German and the English-French retrieval task, the overlap is even lower with 86--87\% pairs having no category overlap. 
These results emphasize the necessity for inexact matching methods such as our learned similarity metric over category embeddings.

\subsection{Patent Prior Art Search}

Cross-lingual patent retrieval is a classic task in CLIR and economically extremely relevant. If a company wants to file a patent application, it is important that the new patent cites all previous patents that are relevant to the claim of its originality. The task of identifying relevant patents is called ``patent prior art search''. In practice, the patent applicant adds all citations that are relevant to the best of his knowledge, and then this list is refined by patent examiners specifically trained on certain areas of technology. 

\subsubsection{Patent Data}

Relevance levels were taken from the publicly available BoostCLIR dataset \cite{SokolovETAL:2013b}. This is a bilingual Japanese-English corpus extracted from the MAREC\footnote{\texttt{http://www.ifs.tuwien.ac.at/imp/marec.shtml}} patent data, and the data from the NTCIR PatentMT\footnote{\texttt{http://research.nii.ac.jp/ntcir/data/data-en.html}} workshop. We make use of all three levels of relevance available: highest relevant patents are family patents (relevance level $r=3$), very relevant patents are the ones cited in search reports by patent examiners ($r=2$), and lowest relevant patents are applicants' added citations ($r=1$). While we use the same distribution of patents for train/dev/test as in BoostCLIR, we changed query and the document content. In the original dataset, both queries and documents consist of patent abstracts. 
However, a patent's technical terms, the scope, and the extent of protection are defined by its claims. 
We thus evaluated a what we think more realistic scenario where the textual part of queries is represented by the patent's title and abstract, and the textual part of documents is represented by the patent's first claim. 

\subsubsection{International Patent Classification}

Meta information for our patent retrieval task consists of the International Patent Classification~(IPC) codes.\footnote{\url{https://www.wipo.int/classifications/ipc/en/}} 
More precisely, we use the sub-group level of the IPC based domestic classification ECLA for English and FI for Japanese.
Our dataset contains 50,329 unique IPC subdivisions for English and 23,020 for Japanese.
We employ the same DeepWalk training strategy to learn classification embeddings as for Wikipedia categories. 
For strictly hierarchical graphs like the IPC-tree, there exist algorithms that better capture hierarchical structures (\newcite{NickelKiela:2017}, \newcite{AlsuhaibaniETAL:2019}, \newcite{LiETAL:2016}, \textit{inter alia}). However, we still chose our previous method for its simplicity and for comparability to previous results. Evaluating truly hierarchical embeddings for patent retrieval is planned as future work. 

\section{Experiments}
\label{experiments}

To evaluate the efficacy of our model, we conduct experiments on Wikipedia data for three language pairs, and on patent data for a single language pair. 
We conduct additional experiments in the patent domain to evaluate alternative models that combine information differently. 
Finally, we integrate standard tf-idf and evaluate if our models scale up to realistic retrieval scenarios in prior art search.

\subsection{Experimental Setup for Wikipedia Retrieval}

We evaluate our model on a cross-lingual retrieval task where we enrich the data provided by \newcite{SasakiETAL:18} with category information from Wikipedia. The textual part of their published data of artificially generated source queries (English) and shortened target documents (Japanese, German, French) remains untouched. 
However, our model integrates the lesser relevant documents ($r=1$) and compares them to the highest relevant documents ($r=2$) and irrelevant documents ($r=0$). We remove all punctuation, and tokenize and lowercase both queries and documents. The same procedure was applied to the Wikipedia corpus consisting of complete articles that we used for unsupervised pre-training of language-dependent word embeddings with \texttt{gensim} \cite{RehurekSoika:10}. 

The statistics of our data are as follows. During training, we use 14.8k documents with $r=2$ and 15.7k with $r=1$ for En-Ja, 32.6k and 34.4k for En-De, and 38.0k and 29.8k for En-Fr, respectively. For practical reasons we do not make use of all available documents with $r=1$ but sample from the set. 
We combine relevant documents with 4 irrelevant documents ($r=0$) for pairwise ranking during training. Depending on the language pair, we evaluate on average 5--8 times more irrelevant than relevant documents per query while testing.
The number of training instances per language pair are 899k, 1.976M, and 2.196M, for En-Ja, En-De, and En-Fr, respectively. The numbers of instances in the dev and test sets are similar, because here we take more irrelevant documents into account (10--15 times more irrelevant than relevant documents).

Our model uses word embeddings of 100 and category embeddings of 30 dimensions. Kernel size for word convolutions is 4, for category convolutions 1. All hyperparameters were tuned on the dev set. The similarity function's MLP implements a 4-layer architecture with 1,600 hidden dimensions where we apply a dropout \cite{SrivastavaETAL:14} rate of 0.5 during training. We trained for 20 epochs with Adam \cite{KingmaBa:14} and selected the best model on the dev set (early stopping). All embeddings were pre-trained and kept fixed during optimization of our ranking model. 

We evaluate our models on three different runs with different random seeds, such that irrelevant ($r=0$) and lesser relevant articles ($r=1$) are varying in the training and dev sets for different runs. For the test set, which always contains all relevant articles, we varied the number of irrelevant articles from 40 to 1,000 to increase the difficulty of the task.

\subsection{Results on Wikipedia}

\begin{table}[t!]
\centering
{\footnotesize
\caption{NDCG results on the Wikipedia retrieval task for ranking with text and text+category embeddings, for language pairs English-Japanese (En-Ja), English-German (En-De), English-French (En-Fr), and against different numbers of irrelevant documents. Significance levels are calculated for corresponding runs (1/2/3), where $^{\dagger}$ denotes $p<0.002$. Our baseline is the ``text only'' model. }
\label{table_wikipedia}
\begin{tabular}{lccccccc}
\toprule
language  & no.irrel. & text only  & text+meta & meta only  & text only  & text+meta & difference  \\
   pair & docs & (run 1/2/3) & (run 1/2/3) & average & average & average & to baseline\\
\midrule

\multirow{3}{*}{\rotatebox[origin=c]{90}{\textbf{En-Ja}}} & $40$ & $89.4/89.4/89.3$ & $90.2^{\dagger}/90.0^{\dagger}/90.0^{\dagger}$ & $66.5$ & $89.4$ & $\bf{90.0}$ & $+0.6$\\
 & $200$ & $78.8/78.2/78.0$ & $79.0^{\dagger}/79.2^{\dagger}/79.3^{\dagger}$ & $51.1$ & $78.3$ & $\bf{79.2}$ & $+0.9$\\
 & $1,000$ & $62.4/61.3/61.2$ & $62.3/62.8^{\dagger}/62.8^{\dagger}$ & $35.9$ & $61.6$ & $\bf{62.6}$ & $+1.0$\\
\midrule
\multirow{3}{*}{\rotatebox[origin=c]{90}{\textbf{En-De}}} & $40$ & $90.4/90.2/90.7$ & $90.9^{\dagger}/90.7^{\dagger}/90.8^{\dagger}$ & $72.0$ & $90.4$ & $\bf{90.8}$ & $+0.4$\\
 & $200$ & $80.2/79.8/80.7$ & $81.7^{\dagger}/81.0^{\dagger}/81.2^{\dagger}$ & $54.8$ & $80.2$ & $\bf{81.3}$ & $+1.1$\\
 & $1,000$ & $64.4/64.0/65.0$ & $67.0^{\dagger}/65.9^{\dagger}/66.3^{\dagger}$ & $ 38.1 $ & $64.5$ & $\bf{66.4}$ & $+1.9$\\
\midrule
\multirow{3}{*}{\rotatebox[origin=c]{90}{\textbf{En-Fr}}} & $40$ & $90.3/90.4/90.4$ & $91.0^{\dagger}/90.9^{\dagger}/91.0^{\dagger}$ & $71.2$ & $90.4$ & $\bf{91.0}$ & $+0.6$\\
 & $200$ & $78.7/79.1/78.7$ & $80.3^{\dagger}/80.2^{\dagger}/80.1^{\dagger}$ & $53.5$ & $78.8$ & $\bf{80.2}$ & $+1.4$\\
 & $1,000$ & $62.1/62.5/62.1$ & $64.3^{\dagger}/64.1^{\dagger}/64.0^{\dagger}$ & $ 36.9 $ & $62.2$ & $\bf{64.1}$ & $+1.9$\\

\bottomrule
\end{tabular}
}

\end{table}

Table \ref{table_wikipedia} shows NDCG \cite{JarvelinKekalainen:02} results for learning-to-rank on a Wikipedia cross-lingual retrieval task for the language pairs English-Japanese (En-Ja), English-German (En-De), English-French (En-Fr). 
We use the original \texttt{trec\_eval} script\footnote{\url{https://trec.nist.gov/trec_eval/}} version 9.07 for evaluation and report standard NDCG without cut-off.
The column labeled \emph{text only} reports results for applying the model of \newcite{SasakiETAL:18} that learns textual convolutions and the similarity function as part of the learning-to-rank task to our multi-level dataset. The best results of our experiments are listed in column \emph{text+meta}. They are obtained by jointly learning the convolutional matrices and the deep layer of the ranking score model (parameters $\vec{O}, \vec{U}$). 
As shown in Table \ref{table_wikipedia}, we achieve significant gains of up to 2 NDCG points averaged over three runs, depending on the language pair and test set sizes. Gains for individual runs were up to 2.6 NDCG points. Significance levels were obtained by running a paired randomization test on corresponding runs as described by \newcite{SmuckerETAL:07}.

Across all language pairs, the integration of category embeddings shows sigificant improvements over the ``text only'' baseline. The absolute improvement is even higher when the difficulty of the task is increased by adding more irrelevant documents to the test set, emphasizing the utility of our pre-trained embeddings for the cross-lingual retrieval task. At the same time, we observe noticeable smaller gains for the English-Japanese pair, which can be explained by the considerably smaller amount of category information in the Japanese version of Wikipedia (172k categories) compared to the French and German version (332k and 282k categories, respectively).

\begin{table}[t]
\centering
\caption{NDCG results on the Japanese-English patent retrieval task for ranking with text and text+meta embeddings, and against different numbers of irrelevant documents. Scores are averaged over six runs. For clarity, we list only the first three runs. 
The model type ``joint'' learns convolutional filters and the similarity function jointly for text and meta data, while the model type ``stacked'' combines the individual scores of ``text only'' and ``meta only'' models linearly. Significance levels are calculated for corresponding runs (1/2/3), where $^{\dagger}$ 
denotes $p<0.0001$. Baseline is the ``text only'' model.}

{\footnotesize

\label{table_patent}
\begin{tabular}{lccccccc}
\toprule
model & no.irrel. & text only  & text+meta & meta only  & text only  & text+meta & difference  \\
type & docs & (run 1/2/3) & (run 1/2/3) & average & average & average & to baseline\\
\midrule
\multirow{4}{*}{\rotatebox[origin=c]{90}{\textbf{joint}}} & $40$ & $92.6/92.8/92.5$ & $92.9/92.8/92.9$ & $ 80.9$ & $92.6$ & $\bf{92.8}$ & $+0.2$\\
 & $200$ & $84.9/85.4/84.9$ & $85.2/85.4/86.9^{\dagger}$ & $ 68.4$ & $85.1$ & $\bf{85.9}$ & $+0.8$\\
 & $400$ & $79.3/81.0/80.0$ & $80.4^{\dagger}/81.0/82.4^{\dagger}$ & $ 62.3$ & $80.1$ & $\bf{81.3}$ & $+1.2$\\
 & $1,000$ & $71.1/72.7/71.4$ & $72.4^{\dagger}/72.6/74.5^{\dagger}$ & $ 53.1$ & $71.9$ & $\bf{73.2}$ & $+1.3$\\
\midrule
\multirow{4}{*}{\rotatebox[origin=c]{90}{\textbf{stacked}}} & $40$ & $92.6/92.8/92.5$ & $93.5^{\dagger}/93.8^{\dagger}/93.6^{\dagger}$ & $ 80.9$ & $92.7$ & $\bf{93.6}$ & $+0.9$\\
 & $200$ & $84.9/85.4/84.9$ & $88.1^{\dagger}/88.5^{\dagger}/87.9^{\dagger}$ & $ 68.4$ & $85.0$ & $\bf{88.2}$ & $+3.2$\\
 & $400$ & $79.3/81.0/80.0$ & $84.1^{\dagger}/85.3^{\dagger}/84.2^{\dagger}$ & $ 62.3$ & $80.1$ & $\bf{84.4}$ & $+4.3$\\
 & $1,000$ & $71.1/72.7/71.4$ & $77.1^{\dagger}/78.4^{\dagger}/77.1^{\dagger}$ & $ 53.1$ & $71.2$ & $\bf{77.5}$ & $+6.3$\\
\bottomrule
\end{tabular}
}
\end{table}

\subsection{Experimental Setup for Patent Search}

For cross-lingual prior art search, we trained a model similar to the Wikipedia model, i.e., embedding dimensions of 100 for text and 30 for meta-information, and a 2-layer similarity network with 3,200 units for the hidden layer. 
IPC codes were extracted from the MAREC corpus, and we apply the same train/dev/test splits as \newcite{SokolovETAL:2013b}. 
During training, we again pair all relevant patents with 4 sampled irrelevant patents to jointly learn the convolutional filters and the similarity function. As before, we evaluate our models on multiple different runs with different random seeds. For the test set, which always contains all relevant articles, we varied the number of irrelevant articles from 40 to 1,000 to increase the difficulty of the task.

Our first experiment evaluates an architecture similar to the one applied on Wikipedia data, while our second experiment combines text and meta-textual models in an ensemble. Following standard practice in CLIR, we conducted a third experiment that selects irrelevant documents based on highest tf-idf scores \cite{Jones:1972} between the Google-translated query and the target document (selection criterion ``tf-idf''). 
After selecting the top-$n$ ($n=$ 40, 200, 400, 1000) irrelevant documents with highest tf-idf score, all relevant documents are added to the list. 
We then applied standard reranking strategy as well as a \textit{weighted} reranking strategy where the final score is a linear combination of tf-idf and ranking score.

\begin{table}[t]
\centering
\caption{NDCG results on the Japanese-English patent retrieval task for ranking with tf-idf, text and text+meta embeddings, and against different numbers of irrelevant documents. Average scores are calculated over six runs. For clarity, we list only the first three runs. 
Here, irrelevant documents are pre-selected per query based on high tf-idf score. 
The ``reranking'' strategy reports numbers by applying the ranking models on the pre-selected list, while the ``weighted reranking'' strategy reports results based on linear combinations of tf-idf and ranking models. 
The penultimate column list the difference to the ``text avg.'' and ``text+tf-idf'' baselines for ``reranking'' and ``weighted reranking'', respectively. The last column lists the difference to plain tf-idf ranking. 
Significance levels are calculated for corresponding runs (1/2/3), where~$^{\dagger}$ denotes $p<10^{-6}$. 
}
\label{table_patent_tfidf}
{\footnotesize
\begin{tabular}{llccccccccc}
\toprule
\multicolumn{2}{l}{strategy} & no.irrel. & tf-idf & text & text+meta & meta & text & text+meta & $\Delta$ to  & $\Delta$ to \\
 & & docs & & (run 1/2/3) & (run 1/2/3) & avg. & avg. & average & text & tf-idf\\
\midrule
\multirow{4}{*}{\rotatebox[origin=c]{90}{\textbf{reranking}}} & &$40$ & $61.9$ & $66.2/66.8/66.1$ & $69.7^{\dagger}/70.2^{\dagger}/69.5^{\dagger}$ & $ 63.6$ & $66.5$ & $\bf{69.8}$ & $+3.3$ & $+7.9$\\
 && $200$ & $ 56.3$ & $51.5/52.4/51.7$ & $55.8^{\dagger}/56.6^{\dagger}/55.5^{\dagger}$ & $ 47.1$ & $52.0$ & $\bf{55.9}$ & $+3.9$ & $-0.4$\\
 && $400$ & $ 55.0$ & $46.1/47.4/46.5$ & $50.8^{\dagger}/51.8^{\dagger}/50.5^{\dagger}$ & $ 41.3$ & $46.8$ & $\bf{50.9}$ & $+4.1$ & $-4.1$\\\
 && $1,000$ & $ 53.9$ & $40.5/41.7/40.8$ & $45.3^{\dagger}/46.2^{\dagger}/45.1^{\dagger}$ & $ 35.1$ & $41.1$ & $\bf{45.5}$ & $+4.4$ & $-8.4$\\
\midrule
 & & & & +tf-idf & +tf-idf & +tf-idf & +tf-idf & +tf-idf & +tf-idf & \\
\midrule
\multirow{4}{*}{\rotatebox[origin=c]{90}{\textbf{weighted}}} & \multirow{4}{*}{\rotatebox[origin=c]{90}{\textbf{reranking}}} & $40$ & $61.9 $ & $66.5/67.2/66.7$ & $69.8^{\dagger}/70.2^{\dagger}/69.5^{\dagger}$ & $ 64.2$ & $66.8$ & $\bf{69.8}$ & $+3.0$ & $+7.9$\\
 && $200$ & $56.3 $ & $60.0/60.1/59.9$ & $61.9^{\dagger}/62.1^{\dagger}/61.7^{\dagger}$ & $ 58.4$ & $60.0$ & $\bf{61.9}$ & $+1.9$ & $+5.6$\\
 && $400$ & $55.0 $ & $58.5/58.8/58.5$ & $60.4^{\dagger}/60.8^{\dagger}/60.5^{\dagger}$ & $ 57.1$ & $58.6$ & $\bf{60.5}$ & $+1.9$ & $+5.5$\\
 && $1,000$ & $53.9 $ & $57.3/57.7/57.2$ & $59.2^{\dagger}/59.5^{\dagger}/59.1^{\dagger}$ & $ 56.1$ & $57.4$ & $\bf{59.2}$ & $+1.8$ & $+5.3$\\
\bottomrule
\end{tabular}
}
\end{table}

\subsection{Results on Patent Data}

\subsubsection{Ranking Models}

Our results listed in the upper half of Table \ref{table_patent} show gains for cross-lingual prior art search similar to our experiments integrating meta-information in CLIR on Wikipedia. The gains on our simplest test set with only 40 irrelevant documents, however, are negligible. This is verified by a paired randomization test \cite{SmuckerETAL:07} to determine significance levels, where none of the three models achieved a p-value below 0.01 when compared to their corresponding ``text only'' systems. Increasing the number of irrelevant documents makes the task harder and with more than 400 irrelevant documents two out of three systems are significantly better than their baseline systems. 

One noticeable difference to the Wikipedia results is the performance of the ``meta only'' system: models solely based on IPC-class embeddings (``meta average'') give much higher retrieval performance than the ones based on Wikipedia category embeddings. We thus evaluated an alternative model we call ``stacked'' (inspired by \newcite{Wolpert:92}), that linearly combines the scores of individual ``text only'' and ``meta only'' models. Our weights were determined by grid search, but they could be in principle optimized using machine learning techniques. 

Interestingly, the combination of text and meta information works far better for the stacked model than for the joint model. We ported back the stacking idea to the Wikipedia retrieval task but observed only minimal changes there. We hypothesize that different orthogonal information sources might not always be captured optimally in a joint model, especially in cases where each single information type already provides a strong signal. In such cases a similarity metric that learns how to optimally compare such information across languages is beneficial. A joint similarity function that is learned on concatenated representations as in the ``joint'' model might blur useful properties of orthogonal information types.

\subsubsection{Scaling-up Retrieval with Weighted Reranking}

Table~\ref{table_patent_tfidf} lists the results of the experiments that integrate tf-idf into the retrieval task. In these experiments, tf-idf is applied to generate subsets of 40, 200, 400, or 1,000 highest scoring documents, and the documents are not randomly sampled. The ranking models are then applied on these subsets. Due to the structure of the original BoostCLIR dataset we could not employ a ranking experiment on the full dataset, but we calculated tf-idf for all 100k+100k
patents listed in the provided dev and test document files of the corpus. 
We then evaluated two retrieval scenarios: in our ``reranking'' approach, we rerank the subset of documents that were selected via tf-idf, while in our ``weighted reranking'' approach a linear combination of tf-idf scores and model scores is applied to this subset. 
In comparison to the previous results, the resulting NDCG scores are considerably lower but reflect a realistic scenario: tf-idf is used to select subsets of top-$k$ ($k=$ 40, 200, 400, or 1,000) highest scoring documents, and then the ranking models are applied to this sets. Combining this strategy with standard IR techniques such as inverted indexing makes our ranking models applicable to large-scale datasets. 

The benefit of including meta information is clearly visible across all subset sizes (``$\Delta$ to text'' in Table~\ref{table_patent_tfidf}). It is important to note that the tf-idf score between the documents and the Google-translated queries is a strong baseline as illustrated in column ``$\Delta$ to tf-idf'' of Table~\ref{table_patent_tfidf}. 
Our experiments that implement a standard reranking strategy, i.e. rerank a pre-selected subset, were only able to surpass plain tf-idf in the one case where the number of documents is rather small, i.e. 40 documents. 

\begin{figure}[t!]
\centering
\subfloat[Reranking]{\includegraphics[width=.35\linewidth]{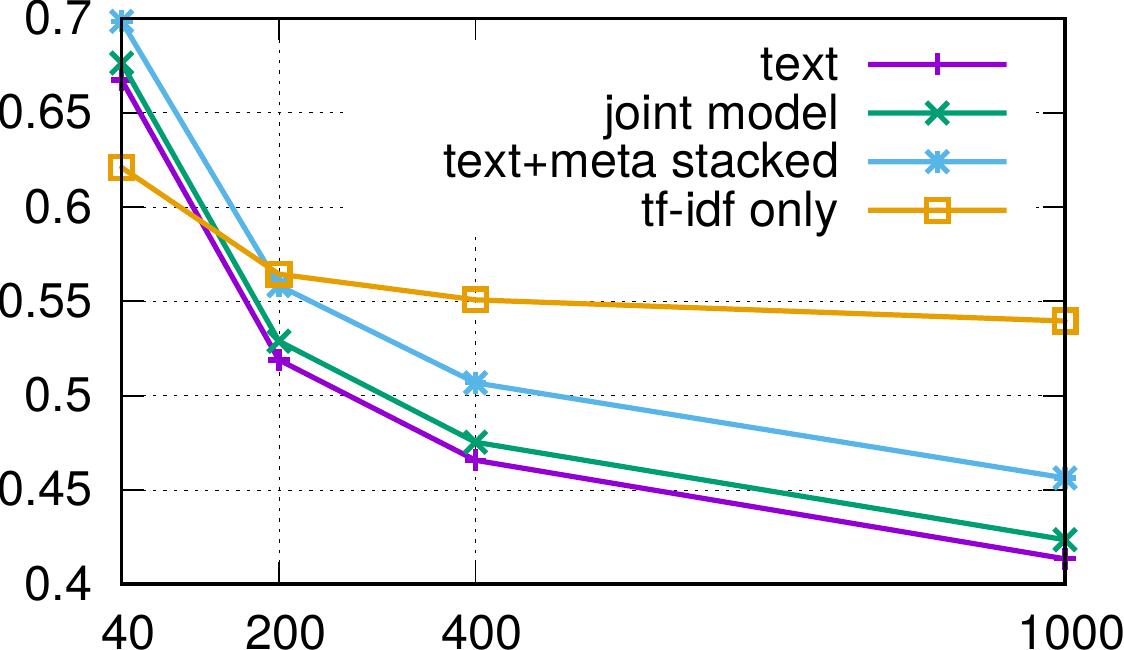}}\quad\quad
\subfloat[Weighted Reranking]{\includegraphics[width=.35\linewidth]{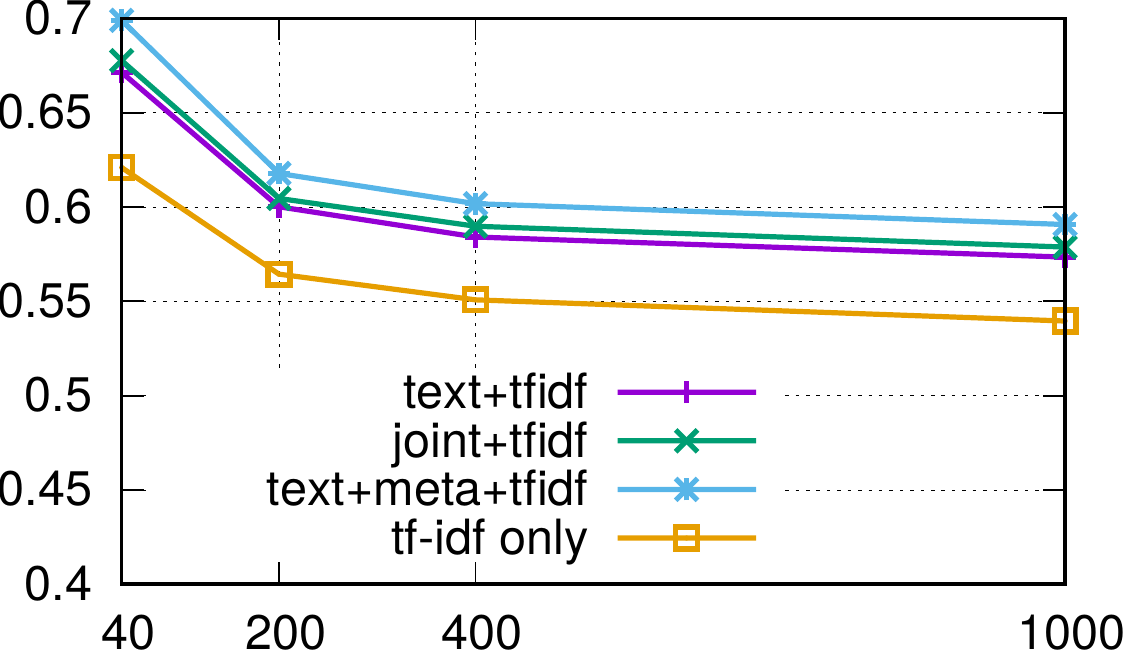}}
\vspace{-2mm}
\caption{NDCG score of reranking and weighted reranking on the dev set. Reranking models (a) are outperformed by plain tf-idf if 200 or more documents are evaluated. Weighted reranking models (b) are able combine different types of information more efficiently and are always above the tf-idf baseline.}
\vspace{-3mm}
\label{fig:reranking_weighted}
\end{figure}

Thus, we reapplied the stacking idea described in the previous section and evaluated combinations of model scores and tf-idf scores. For all combinations, the weights were determined via grid search on dev (Figure~\ref{fig:reranking_weighted}). Again, we observe remarkable gains in retrieval performance when models that each provide strong signals are ``stacked'' together. This time, the tf-idf baseline is outperformed in all cases by a large margin of 5.3--7.9 NDCG points. 
The positive contribution of integrating meta information is not as large as in the standard reranking experiment, however, it is still clearly evident on this relatively high level.

\section{Conclusion}

We presented an approach to incorporate retrieval-relevant meta-textual information into learning-to-rank models. Such information is readily available in many retrieval tasks and can be integrated by learning dense embeddings that allow for inexact matching. 
Our main motivation was to investigate the relative benefit of incorporating meta information into a manageable neural architecture presented in previous work \cite{SasakiETAL:18}. 
If more advanced and potentially better performing embedding models like BERT \cite{DevlinETAL:2018} or ELMo \cite{PetersETAL:2018} can be further enhanced by meta-textual information is an open question for future work.

Our results in the Wikipedia domain show that adding category information yields significant gains across several language pairs. The contribution of meta information increases with the size of the documents' lists, but is also dependent on the amount of meta information. 
In the patent domain, adding patent classification embeddings shows significant improvements over the textual models. Existing models that make use of the tf-idf metric can be further improved if the system components are combined properly. Integrating the tf-idf metric also makes the ranking models applicable to large scale retrieval if a machine translation system is available to translate queries into the target language.

Finally, our experiments on two different data sets and up to three language pairs showed that 
combining different types of information in retrieval 
is not straightforward. Joint learning worked particularly well for integrating noisy Wikipedia categories, while for the cleaner hierarchical patent IPC classifications we observed the largest gains when individual models are 
are combined in an ensemble setup. On patent data, joint models were significantly better than single models, but weighted ensembles outperformed joint models by a large margin.

\bibliographystyle{coling}
\bibliography{references}

\begin{thebibliography}{}

\bibitem[\protect\citename{Alsuhaibani \bgroup et al.\egroup
  }2019]{AlsuhaibaniETAL:2019}
Mohammed Alsuhaibani, Takanori Maehara, and Danushka Bollegala.
\newblock 2019.
\newblock Joint learning of hierarchical word embeddings from a corpus and a
  taxonomy.
\newblock In {\em 1st Conference on Automated Knowledge Base Construction,
  {AKBC} 2019}, Amherst, MA.

\bibitem[\protect\citename{Chen \bgroup et al.\egroup }2016]{ChenETAL:16}
Wenhu Chen, Evgeny Matusov, Shahram Khadivi, and Jan-Thorsten Peter.
\newblock 2016.
\newblock Guided alignment training for topic-aware neural machine translation.
\newblock In {\em Proceedings of the 12th Conference of the Association for
  Machine Translation in the Americas {(AMTA)}}, Austin, TX.

\bibitem[\protect\citename{Chu \bgroup et al.\egroup }2017]{ChuETAL:17}
Chenhui Chu, Raj Dabre, and Sadao Kurohashi.
\newblock 2017.
\newblock An empirical comparison of domain adaptation methods for neural
  machine translation.
\newblock In {\em Proceedings of the 55th Annual Meeting of the Association for
  Computational Linguistics (Volume 2: Short Papers)}, Vancouver, Canada.

\bibitem[\protect\citename{Devlin \bgroup et al.\egroup }2019]{DevlinETAL:2018}
Jacob Devlin, Ming{-}Wei Chang, Kenton Lee, and Kristina Toutanova.
\newblock 2019.
\newblock {BERT:} pre-training of deep bidirectional transformers for language
  understanding.
\newblock In {\em Proceedings of the 2019 Conference of the North American
  Chapter of the Association for Computational Linguistics: Human Language
  Technologies, {NAACL-HLT} 2019}, Minneapolis, MN.

\bibitem[\protect\citename{Grover and Leskovec}2016]{GroverLeskovec:2016}
Aditya Grover and Jure Leskovec.
\newblock 2016.
\newblock node2vec: Scalable feature learning for networks.
\newblock In {\em Proceedings of the 22nd {ACM} {SIGKDD} International
  Conference on Knowledge Discovery and Data Mining}, San Francisco, CA.

\bibitem[\protect\citename{J\"arvelin and
  Kek\"al\"ainen}2002]{JarvelinKekalainen:02}
Kalervo J\"arvelin and Jaana Kek\"al\"ainen.
\newblock 2002.
\newblock Cumulated gain-based evaluation of {IR} techniques.
\newblock {\em {ACM} Transactions in Information Systems}, 20(4):422--446.

\bibitem[\protect\citename{Jehl and Riezler}2018]{JehlRiezler:18}
Laura Jehl and Stefan Riezler.
\newblock 2018.
\newblock Document information as side constraints for improved patent
  translation.
\newblock In {\em Proceedings of the 13th biennial conference of the
  Association for Machine Translation in the Americas {(AMTA)}}, Boston, {MA},
  {USA}.

\bibitem[\protect\citename{Johnson \bgroup et al.\egroup }2017]{JohnsonETAL:17}
Melvin Johnson, Mike Schuster, Quoc~V Le, Maxim Krikun, Yonghui Wu, Zhifeng
  Chen, Nikhil Thorat, Fernanda Vi{\'e}gas, Martin Wattenberg, Greg Corrado,
  et~al.
\newblock 2017.
\newblock Google's multilingual neural machine translation system: enabling
  zero-shot translation.
\newblock {\em Transactions of the Association for Computational Linguistics
  {(TACL)}}, 5:339--351.

\bibitem[\protect\citename{Jones}1972]{Jones:1972}
Karen~Spärck Jones.
\newblock 1972.
\newblock A statistical interpretation of term specificity and its application
  in retrieval.
\newblock {\em Journal of Documentation}, 28:11--21.

\bibitem[\protect\citename{Kingma and Ba}2015]{KingmaBa:14}
Diederik~P. Kingma and Jimmy Ba.
\newblock 2015.
\newblock Adam: {A} method for stochastic optimization.
\newblock In {\em 3rd International Conference on Learning Representations,
  {ICLR} 2015}, San Diego, CA.

\bibitem[\protect\citename{Kobus \bgroup et al.\egroup }2017]{KobusETAL:17}
Catherine Kobus, Josep Crego, and Jean Senellart.
\newblock 2017.
\newblock Domain control for neural machine translation.
\newblock In {\em Proceedings of Recent Advances in Natural Language Processing
  {(RANLP)}}, Varna, Bulgaria.

\bibitem[\protect\citename{Li \bgroup et al.\egroup }2016]{LiETAL:2016}
Yuezhang Li, Ronghuo Zheng, Tian Tian, Zhiting Hu, Rahul Iyer, and Katia~P.
  Sycara.
\newblock 2016.
\newblock Joint embedding of hierarchical categories and entities for concept
  categorization and dataless classification.
\newblock In {\em Proceedings of the 26th International Conference on
  Computational Linguistics, {COLING} 2016}, Osaka, Japan.

\bibitem[\protect\citename{Mikolov \bgroup et al.\egroup }2013]{MikolovETAL:13}
Tomas Mikolov, Ilya Sutskever, Kai Chen, Greg~S. Corrado, and Jeff Dean.
\newblock 2013.
\newblock Distributed representations of words and phrases and their
  compositionality.
\newblock In {\em Neural Information Processing Systems {(NIPS)}}, Lake Tahoe,
  {CA}.

\bibitem[\protect\citename{Mitra and Craswell}2018]{MitraCrasswell:18}
Bhaskar Mitra and Nick Craswell.
\newblock 2018.
\newblock An introduction to neural information retrieval.
\newblock {\em Foundations and Trends in Information Retrieval}, 13(1):1--126.

\bibitem[\protect\citename{Nickel and Kiela}2017]{NickelKiela:2017}
Maximilian Nickel and Douwe Kiela.
\newblock 2017.
\newblock Poincar{\'{e}} embeddings for learning hierarchical representations.
\newblock In {\em Advances in Neural Information Processing Systems 30: Annual
  Conference on Neural Information Processing Systems (NIPS)}, Long Beach, CA.

\bibitem[\protect\citename{Perozzi \bgroup et al.\egroup }2014]{PerozziETAL:14}
Bryan Perozzi, Rami Al-Rfou, and Steven Skiena.
\newblock 2014.
\newblock Deepwalk: Online learning of social representations.
\newblock In {\em Proceedings of the 20th ACM SIGKDD International Conference
  on Knowledge Discovery and Data Mining (KDD '14)}, New York, NY.

\bibitem[\protect\citename{Peters \bgroup et al.\egroup }2018]{PetersETAL:2018}
Matthew~E. Peters, Mark Neumann, Mohit Iyyer, Matt Gardner, Christopher Clark,
  Kenton Lee, and Luke Zettlemoyer.
\newblock 2018.
\newblock Deep contextualized word representations.
\newblock In {\em Proceedings of the 2018 Conference of the North American
  Chapter of the Association for Computational Linguistics: Human Language
  Technologies, {NAACL-HLT} 2018}, New Orleans, LA.

\bibitem[\protect\citename{{\v R}eh{\r u}{\v r}ek and
  Sojka}2010]{RehurekSoika:10}
Radim {\v R}eh{\r u}{\v r}ek and Petr Sojka.
\newblock 2010.
\newblock {Software Framework for Topic Modelling with Large Corpora}.
\newblock In {\em {Proceedings of the LREC 2010 Workshop on New Challenges for
  NLP Frameworks}}, Valletta, Malta. ELRA.

\bibitem[\protect\citename{Ribeiro \bgroup et al.\egroup
  }2017]{FigueiredoETAL:2017}
Leonardo Filipe~Rodrigues Ribeiro, Pedro H.~P. Saverese, and Daniel~R.
  Figueiredo.
\newblock 2017.
\newblock struc2vec: Learning node representations from structural identity.
\newblock In {\em Proceedings of the 23rd {ACM} {SIGKDD} International
  Conference on Knowledge Discovery and Data Mining}, Halifax, NS, Canada.

\bibitem[\protect\citename{Sasaki \bgroup et al.\egroup }2018]{SasakiETAL:18}
Shota Sasaki, Shuo Sun, Shigehiko Schamoni, Kevin Duh, and Kentaro Inui.
\newblock 2018.
\newblock Cross-lingual learning-to-rank with shared representations.
\newblock In {\em Proceedings of the Conference of the North American Chapter
  of the Association for Computational Linguistics: Human Language Technologies
  {(NAACL-HLT)}}, New Orleans, {LA}.

\bibitem[\protect\citename{Schamoni and Riezler}2015]{SchamoniRiezler:15}
Shigehiko Schamoni and Stefan Riezler.
\newblock 2015.
\newblock Combining orthogonal information in large-scale cross-language
  information retrieval.
\newblock In {\em Proceedings of the 38th Annual {ACM SIGIR} Conference},
  Santiago, Chile.

\bibitem[\protect\citename{Sennrich and Haddow}2016]{SennrichETAL16linguistic}
Rico Sennrich and Barry Haddow.
\newblock 2016.
\newblock {Linguistic Input Features Improve Neural Machine Translation}.
\newblock In {\em Proceedings of the First Conference on Machine Translation
  (WMT)}, Berlin, Germany.

\bibitem[\protect\citename{Sennrich \bgroup et al.\egroup
  }2016]{SennrichETAL:16}
Rico Sennrich, Barry Haddow, and Alexandra Birch.
\newblock 2016.
\newblock {Controlling Politeness in Neural Machine Translation via Side
  Constraints}.
\newblock In {\em Proceedings of the Conference of the North American Chapter
  of the Association for Computational Linguistics: Human Language Technologies
  {(NAACL-HLT)}}, San Diego, {CA}.

\bibitem[\protect\citename{Smucker \bgroup et al.\egroup }2007]{SmuckerETAL:07}
Mark~D. Smucker, James Allan, and Ben Carterette.
\newblock 2007.
\newblock A comparison of statistical significance tests for information
  retrieval evaluation.
\newblock In {\em CIKM '07: Proceedings of the sixteenth ACM conference on
  Conference on information and knowledge management}, New York, NY.

\bibitem[\protect\citename{Sokolov \bgroup et al.\egroup
  }2013]{SokolovETAL:2013b}
Artem Sokolov, Laura Jehl, Felix Hieber, and Stefan Riezler.
\newblock 2013.
\newblock Boosting cross-language retrieval by learning bilingual phrase
  associations from relevance rankings.
\newblock In {\em Proceedings of the Conference on Empirical Methods in Natural
  Language Processing (EMNLP)}, Seattle, WA.

\bibitem[\protect\citename{Srivastava \bgroup et al.\egroup
  }2014]{SrivastavaETAL:14}
Nitish Srivastava, Geoffrey Hinton, Alex Krizhevsky, Ilya Sutskever, and Ruslan
  Salakhutdinov.
\newblock 2014.
\newblock Dropout: A simple way to prevent neural networks from overfitting.
\newblock {\em Journal of Machine Learning Research}, 15:1929--1958.

\bibitem[\protect\citename{Wang \bgroup et al.\egroup }2018]{WangETAL:2017}
Hongwei Wang, Jia Wang, Jialin Wang, Miao Zhao, Weinan Zhang, Fuzheng Zhang,
  Xing Xie, and Minyi Guo.
\newblock 2018.
\newblock Graphgan: Graph representation learning with generative adversarial
  nets.
\newblock In {\em Proceedings of the Thirty-Second {AAAI} Conference on
  Artificial Intelligence, (AAAI-18), the 30th innovative Applications of
  Artificial Intelligence (IAAI-18), and the 8th {AAAI} Symposium on
  Educational Advances in Artificial Intelligence (EAAI-18)}, New Orleans, LA.

\bibitem[\protect\citename{Wolpert}1992]{Wolpert:92}
David~H. Wolpert.
\newblock 1992.
\newblock Stacked generalization.
\newblock {\em Neural Networks}, 5(2):241--259.

\end{thebibliography}

\end{document}